\documentstyle [12pt,epsfig]{article} 
\makeatletter

\@addtoreset{equation}{section}
\makeatother

\newcommand{\ba}{\begin{eqnarray}}
\newcommand{\ea}{\end{eqnarray}}

\begin{document}
\newcommand{\BS}{\bigskip}
\newcommand{\SECTION}[1]{\BS{\large\section{\bf #1}}}
\newcommand{\SUBSECTION}[1]{\BS{\large\subsection{\bf #1}}}
\newcommand{\SUBSUBSECTION}[1]{\BS{\large\subsubsection{\bf #1}}}

\begin{titlepage}
\begin{center}
\vspace*{2cm}
{\large \bf Primary and reciprocal space-time experiments, relativistic
  reciprocity relations and Einstein's train-embankment thought experiment}  
\vspace*{1.5cm}
\end{center}
\begin{center}
{\bf J.H.Field }
\end{center}
\begin{center}
{ 
D\'{e}partement de Physique Nucl\'{e}aire et Corpusculaire
 Universit\'{e} de Gen\`{e}ve . 24, quai Ernest-Ansermet
 CH-1211 Gen\`{e}ve 4.
}
\newline
\newline
   E-mail: john.field@cern.ch
\end{center}
\vspace*{2cm}
\begin{abstract}
   The concepts of primary and reciprocal experiments and base and travelling
   frames in special relativity are concisely described and applied to several different
    space-time experiments. These include Einstein's train/embankment thought experiment
    and a related thought experiment, due to Sartori, involving two trains in parallel motion
     with different speeds. Spatially separated clocks which are synchronised in their common
    proper frame are shown to be so in all inertial frames and their spatial separation
    to be Lorentz invariant. The interpretions given by Einstein and Sartori of their
    experiments, as well as those given by the present author in previous
     papers, are shown to be erroneous. 


\vspace*{1cm}
\end{abstract}
\end{titlepage}
 
\SECTION{\bf{Introduction}}
    The aim of the present paper is to present a concise summary and some applications
    of a nomenclature and notation for the general description of space-time 
      experiments introduced and explained in detail in Ref.~\cite{JHFSTP3}.   
     The latter is the third in a series of recently written papers
     ~\cite{JHFSTP1,JHFSTP2,JHFSTP3}, devoted to space time physics in the absence
    of gravitation, that correct several misconceptions about the subject
     originating in Einstein's seminal Special Relativity paper~\cite{Ein1}.
      The most important of these, the spurious nature of the `relativity of
      simultaneity' (RS) and `length contraction' (LC) effects was explained in
      Ref.~\cite{JHFLLT} and further discussed from
      different points-of-view in Refs.~\cite{JHFCRCS,JHFACOORD,JHFUMC}.
      At the time of writing, there is ample and precise experimental confirmation
     of the time dilation (TD) effect, predicted as a consequence of the space-time
      Lorentz Transformation (LT) in Ref.~\cite{Ein1}, but none for RS or
      LC~\cite{JHFLLT}. Earth-satellite based experiments to test for the existence
     of RS have been proposed by the present author~\cite{JHFSEXPS}.
      \par The present paper contains, in the following section, definitions
        of the concepts of base and travelling frames and a space-time 
       experiment and its reciprocal, introduced in Ref.~\cite{JHFSTP3}.
       The following sections contain applications of these concepts:
       time dilation and the simultaneity of spatially separated events
     in different inertial frames, the Lorentz  invariance of spatial intervals,
     velocity transformation formulas and reciprocity relations,
    Einstein's train-embankment experiment~\cite{EinTETE}
    and a thought experiment involving two trains moving on parallel tracks
    at different speeds due to Sartori~\cite{Sartori}. As explained below,
    these two thought experiments were incorrectly analysed in previous 
   papers by the present author.

\SECTION{\bf{Base and Travelling frames; Primary and Reciprocal space-time experiments}}

  An experiment is considered where a ponderable physical object at some fixed position in
  an inertial frame, S', is in uniform motion relative to another inertial frame S. The
  frame S is denoted as the {\it base frame} of the experiment, S' as a {\it travelling
  frame}. As is conventional the origin of S' moves along the positive $x$-axis in S with
  speed $v_B$, the $x$- and $x'$-axes being parallel, and the object lying on the $x'$ axis.
  The above configuration describes a {\it primary experiment}; the value of $v_B$ is a fixed
  initial condition specified in the frame S. An experiment with a {\it reciprocal configuration}
   is one in which the origin of S moves along the negative $x'$-axis with speed $v'_B$. S' is 
  now the base frame and S the travelling frame. The value of $v'_B$ is a fixed initial
  condition (in general not equal to $v_B$) specified in the frame S'. In the special case
   that $v_B = v'_B \equiv v$, the experiment with the reciprocal configuration is termed
   {\it reciprocal} to the primary experiment, and {\it vice versa}.

 \SECTION{\bf{Time dilation and invariance of simultaneity}}
   The nomenclature introduced above is now applied to a primary experiment and its reciprocal,
   in which similar clocks C, C' are situated at the origins of S and S' respectively. In the primary
   experiment C' moves with speed $v_B = v$ along the positive $x$-axis in S, and in the reciprocal
   experiment C moves with speed $v'_B = v$ along the negative $x'$-axis in S'.
    The Lorentz transformations
    (and their inverses) describing the experiments are as follows:
   \par \underline{Primary Experiment}
     \par Transformation:
       \begin{eqnarray}
    x'({\rm C'})_T & = & \gamma[x({\rm C'})_B-vt({\rm C})_B] = 0,~\rightarrow x({\rm C'})_B = vt({\rm C})_B,  \\
    t'({\rm C'})_T & = & \gamma[ t({\rm C})_B- \frac {v x({\rm C'})_B}{c^2}],~\rightarrow t'({\rm C'})_T
      = \frac{t({\rm C})_B}{\gamma}.
     \end{eqnarray}
 \par Inverse Transformation:
      \begin{eqnarray}
    x({\rm C'})_B & = & \gamma[x'({\rm C'})_T+vt'({\rm C'})_T],~\rightarrow x({\rm C'})_B = \gamma vt'({\rm C'})_T = vt({\rm C})_B,  \\
    t({\rm C})_B & = & \gamma[ t'({\rm C'})_T+ \frac {v x'({\rm C'})_T}{c^2}],~\rightarrow t({\rm C})_B
      =  \gamma t'({\rm C'})_T.
     \end{eqnarray}
   \par \underline{Reciprocal Experiment}
     \par Transformation:
       \begin{eqnarray}
    x({\rm C})_T & = & \gamma[x'({\rm C})_B+vt'({\rm C'})_B] = 0,~\rightarrow x'({\rm C})_B = -vt'({\rm C'})_B,  \\
    t({\rm C})_T & = & \gamma[ t'({\rm C'})_B+ \frac {v x'({\rm C})_B}{c^2}],~\rightarrow t({\rm C})_T
      = \frac{t'({\rm C'})_B}{\gamma}.
     \end{eqnarray}
 \par Inverse Transformation:
      \begin{eqnarray}
    x'({\rm C})_B & = & \gamma[x({\rm C})_T-vt({\rm C})_T],~\rightarrow x'({\rm C})_B = -\gamma vt({\rm C})_T =
   -vt'({\rm C'})_B,  \\
    t'({\rm C'})_B & = & \gamma[ t({\rm C})_T-\frac {v x({\rm C})_T}{c^2}],~\rightarrow t'({\rm C'})_B
      =  \gamma t({\rm C})_T.
     \end{eqnarray}
      where $\gamma \equiv 1/\sqrt{1-(v/c)^2}$. $t({\rm C})$ and $t'({\rm C'})$ are the times
     recorded by C and C' respectively and the subscripts $B$ and $T$ specify whether the space
     or time coordinate is defined in a base frame or a travelling frame, respectively. Thus 
     $t({\rm C})_B$ and $t'({\rm C'})_B$ are times recorded by clocks at rest in primary 
     and reciprocal experiments, respectively while $t'({\rm C'})_T$ and $t({\rm C})_T$ are the
     respective times recorded by clocks in motion in the two experiments.The time dilation (TD)
     relations given by the second equations in (3.2), (3.4), (3.6) and (3.8) are obtained by
       using the equations of motion in (3.1), (3.3), (3.5) and (3.7) respectively to eliminate
     the spatial coordinates on the right sides of the first equations in (3.2), (3.4), 
    (3.6) and (3.8).   
     \par The following remarks may be made concerning Eq.~s.(3.1)-(3.8)
    \begin{itemize}
      \item[(i)] The primary experiment and its reciprocal are physically independent.
                 The LT equations for the primary experiment contain only the spatial cordinates
                of the travelling clock C', the position of the stationary base-frame clock C being arbitary.
                The LT equations for the reciprocal experiment contain only the spatial cordinates
                of the travelling clock C, the position of the stationary base frame clock C' being arbitary.
      \item[(ii)] In both experiments, clocks in the travelling (base) frame appear to be running
                  slower (faster) to observers in the base (travelling) frames.
 
       \item[(iii)] Identical predictions are given, in both the primary and reciprocal
                 experiments, by the transformation and the inverse 
                  transformation.

     \item[(iv)] The TD relations:
               \[ t({\rm C})_B  = \gamma t'({\rm C'})_T;~~~t'({\rm C'})_B =  \gamma t({\rm C})_T \]
                 are translationally invariant (do not depend on the spatial positions of the clocks).
    \item[(v)]  The equations of motion of the clocks:
                  \[ x'({\rm C'})_T = 0,~~~ x({\rm C'})_B = vt({\rm C})_B;~~ x({\rm C})_T = 0,~~~ x'({\rm C})_B = -vt'({\rm C'})_B \]
                 are the same as in Galilean relativity.
      \end{itemize}
       Because of (iv) the TD relations hold for pairs of clocks, at arbitary positions in S and S';
         that is: 
          \begin{eqnarray}
  t({\rm C_1})_B  & = & \gamma t'({\rm C'_1})_T,  \\
t({\rm C_2})_B  & = & \gamma t'({\rm C'_2})_T 
     \end{eqnarray}
   where C$_1$ and C$_2$ are at arbitary positions in S and   C$'_1$ and C$'_2$ are at arbitary
   positions in S. If now C$'_1$ and C$'_2$ are synchronised so that, at any instant in the frame S':
       \begin{equation}
        t'({\rm C'_1})_T = t'({\rm C'_2})_T = t'_T 
  \end{equation}
  it follows from (3.9) and (3.10) that:
         \begin{equation}
        t({\rm C_1})_B = t({\rm C_2})_B= \gamma t'_T = t_B. 
  \end{equation}   
      There is therefore no `relativity of simultaneity' effect for a pair of synchronised clocks
     at different positions in S´  ---they are also observed to be synchronised in the frame S.
      How this spurious effect arises from misuse of the space-time Lorentz transformation
       is explained elsewhere~\cite{JHFSTP1,JHFSTP2,JHFSTP3,JHFLLT,JHFCRCS,JHFACOORD,JHFUMC}.

      \SECTION{\bf{Lorentz invariance of spatial separations}}
          To discuss spatial intervals using the Lorentz transformation, an abbreviated notation
         is used where the clock at the origin of S' with $x'_T =0$  is given the label 1 and a second clock,
         on the $x'$ axis, with $x'_T = L'$ the label 2. Assuming the same initial conditions for the primary
         experiment as in Eq.~s(3.1) and (3.2) and dropping, for simplicity, the clock, base frame and travelling
         frame labels, Eq.~s(3.1) and (3.2) are written:
         \begin{eqnarray}
          x'_1  & = &\gamma(x_1-vt_1) = 0,~~\rightarrow~x_1 = vt_1,   \\
          t'_1 & = &\gamma(t_1-\frac{v x_1}{c^2})~~\rightarrow~ t'_1 =\frac{t_1}{\gamma}.
         \end{eqnarray}
       The equation of motion in S of the clock at  $x'_T =L'$ is
         \begin{equation}
           x_2 = vt_2 + L
         \end{equation}
         where $L \equiv x_2(t_2 = 0)$ is a constant, independent of the value of $v$, depending
         on the choice of spatial coordinates in S. The space transformation equation for the 
         clock 2, consistent
         with (4.1) in the limit $L = L' =0$, and therefore using the same spatial coordinate
        system in S as clock 1, is:
    \begin{equation}
    x'_2-L' =  \gamma(x_2-L-vt_2) = 0,~~\rightarrow~x_2 = vt_2+L. 
      \end{equation}
     The corresponding time transformation equation, given by the replacement $x \rightarrow x-L$, 
     in (4.2) is 
     \begin{equation}
   t'_2  = \gamma[t_2-\frac{v(x_2-L)}{c^2}]~~\rightarrow~ t'_2 =\frac{t_2}{\gamma}.
    \end{equation}
     Considering now simultaneous events in the frame S'; $t'_1 = t'_2 = t'$, (4.1)-(4.5)
     yield:
     \begin{eqnarray}
      x_1(\beta) & = & \beta c t_1 = \gamma \beta c t',   \\
      t_1(\beta)& = & \gamma t', \\
   x_2(\beta)- L & = & \beta c t_2 = \gamma \beta c t',   \\
      t_2(\beta) & = & \gamma t'
     \end{eqnarray}
      where $\beta \equiv v/c$, and the $\beta$ dependences of $x$ and $t$, for a fixed value of
       $t'$, are explicitly indicated.
      \par With the aid of the identity: $\gamma^2 -\gamma^2 \beta^2 \equiv 1$, (4.6),(4.7) and
       (4.8),(4.9) yield identically-shaped hyperbolic curves on the $ct$ versus $x$ plot for a given value of $t'$:
   \begin{equation}
     c^2 t_1(\beta)^2-x_1(\beta)^2 = c^2 (t')^2 =  c^2 t_2(\beta)^2-(x_2(\beta)-L)^2.
   \end{equation}
     Since (4.7) and (4.9) give
      \begin{equation}
   t_1(\beta) = \gamma t'  =  t_2(\beta)
      \end{equation}
    (4.10) simplifies to
   \begin{equation}
    x_2(\beta)- x_1(\beta) = L.
      \end{equation}
    The spatial separation of the clocks in S is therefore independent
   of the value of $\beta$. Since, for $\beta \rightarrow 0$,  $x \rightarrow x'$ it follows
   from (4.12) that:
  \begin{equation}
    x_2(0)- x_1(0) = x'_2- x'_1 \equiv L' =  L.
      \end{equation}
      The spatial separation of the clocks in S and S' is therefore the same for all values
      of $\beta$ ---there is no `relativistic length contraction'. How the latter spurious
      effect ---correlated with `relativity of simultaneity' --- arises is also discussed in
     Refs.~\cite{JHFSTP1,JHFSTP2,JHFSTP3,JHFLLT,JHFCRCS,JHFACOORD,JHFUMC}. 
´
 \SECTION{\bf{Velocity transformation formulas and relativistic reciprocity relations}}
     Two, physically distinct, kinds of velocity addition formulas are considered in this section.
      The first, corresponding to the well-known relativistic velocity addition formulas as derived
      by Einstein in Ref.~\cite{Ein1}, gives relations between the base frame velocities of a single
       object in different inertial frames. The second gives the transformation of the relative
       velocity of two objects in a given inertial frame into the similarly defined relative velocity
      between them in another inertial frame. For the first type of transformation, since only base
      frame velocities are involved, the `travelling frame' concept plays no role, whereas it is
       essential for the second (relative velocity) transformation in order to correctly understand
      the physical basis of the TD effect.
      \par Suppose that the frame S' moves with speed $v_B = v$ in the positive $x$-direction in S and
       that an object moves with velocity components $u^{(x)}_B$ and $u^{(y)}_B$ in the directions 
       of the $x$- and $y$-axes in S. The first type of calculation predicts the corresponding base
       frame velocities  $\bar{w}^{(x')}_B$ and $\bar{w}^{(y')}_B$ in the frame S'. The bar on a symbol
       denotes that it is a derived quantity rather than an assumed initial value of
       a parameter of the problem. The appropriate differential LT formulas are:
       \begin{eqnarray}
        dx'_B & = & \gamma[dx_B - v dt_B], \\
        dy'_B & = & dy_B,  \\
       dt'_B & = & \gamma[dt_B - \frac{v dx_B}{c^2}]
       \end{eqnarray}
        where
      \begin{eqnarray}
       \frac{dx_B}{dt_B} & \equiv  & u^{(x)}_B~,~~~ \frac{dy_B}{dt_B} \equiv u^{(y)}_B, \\
       \frac{dx'_B}{dt'_B} & \equiv  & \bar{w}^{(x')}_B~,~~~ \frac{dy'_B}{dt'_B} \equiv  \bar{w}^{(y')}_B.
      \end{eqnarray}
       Dividing (5.1) or (5.2) by (5.3) and subsituting, in the equations so obtained, the base frame
      velocities defined in (5.4) and (5.5) gives the longitudinal and transverse base frame velocity
      addition formulas:
      \begin{eqnarray}
      \bar{w}^{(x')}_B & = & \frac{u^{(x)}_B-v_B}{1-\frac{v_B u^{(x)}_B}{c^2}},   \\
       \bar{w}^{(y')}_B & = & \frac{u^{(y)}_B}{\gamma(1-\frac{v_B u^{(x)}_B}{c^2})}.
   \end{eqnarray}
        Eqs(5.1)-(5.3) can also be used to derive transformation equations for the 4-vector 
       velocity, $U$, of the object. If $d \tau$ denotes an interval of the proper time of the object,
      the TD relations $dt_B = \gamma_{u_B}d \tau$ and  $dt'_B = \gamma_{\bar{w}_B}d \tau$ 
      ($\gamma_u \equiv 1/\sqrt{1-(u/c)^2}$), where $u_B \equiv \sqrt{(u^{(x)}_B)^2 +(u^{(y)}_B)^2}$, give, on dividing (5.1)-(5.3) throughout by  $d \tau$
       and using (5.4) and (5.5), the relations:
   \begin{eqnarray}
        \gamma_{\bar{w}_B}\bar{w}^{(x')}_B & = & \gamma[\gamma_{u_B} u^{(x)}_B -v \gamma_{u_B}], \\
          \gamma_{\bar{w}_B}\bar{w}^{(y')}_B & = &  \gamma_{u_B} u^{(y)}_B,   \\
   \gamma_{\bar{w}_B} & = & \gamma[\gamma_{u_B} - \frac{ v \gamma_{u_B} u^{(x)}_B}{c^2} ] 
     \end{eqnarray}
      or
      \begin{eqnarray}
          U'^{(x')} & = & \gamma[ U^{(x)} -\beta  U^{(0)}],\\
           U'^{(y')} & = &  U^{(y)},   \\
  U'^{(0)} & = & \gamma[ U^{(0)} -\beta  U^{(x)}]
     \end{eqnarray} 
      where the 4-vector velocities $U$ and $U'$ of the object in S and S' are defined as:
   \begin{eqnarray}
  U & = & (U^{(0)};U^{(x)},U^{(y)},U^{(z)}) \equiv ( c\gamma_{u_B};
  \gamma_{u_B}u^{(x)}_B,\gamma_{u_B}u^{(y)}_B, 0),  \\
  U' & = & (U'^{(0)};U'^{(x')},U'^{(y')},U'^{(z')}) \equiv ( c \gamma_{\bar{w}_B};
      \gamma_{\bar{w}_B} \bar{w}^{(x')},\gamma_{\bar{w}_B} \bar{w}^{(y')}, 0).
  \end{eqnarray} 
  The velocity addition relations (5.6) and (5.7) are recovered by dividing (5.8) and (5.9) respectively
   by (5.10) or dividing (5.11) and (5.12) respectively by (5.13) and using the definitions of the
   components of the 4-vector velocities in (5.14) and (5.15).
    \par It is interesting to note that, although space-time events in an experiment and its reciprocal
    are physically independent, the initial kinematical configurations in the
    two experiments are related by the kinematical LT (5.11)-(5.13) that yields the parallel
     velocity addition formula when $u^{(x)}_B = u_B$ and  $u^{(y)}_B = 0$:
     \begin{equation}
       \bar{w}_B   =   \frac{u_B-v_B}{1-\frac{v_B u_B}{c^2}}.
      \end{equation}
     If $u_B = 0$ (for example an object at rest at the origin of S) than (5.16) gives 
      $-\bar{w}_B = v'_B = v_B$, which describes the kinematical configuration of the reciprocal 
      experiment ---the object moves with speed $v_B$ along the negative $x'$-axis in S'.        
  
    \par Consider now an experiment in which an object moving with the specified speed $u_B$  along the
      positive $x$-axis in S is observed in the travelling frame S'. Since the relative velocity
      of the object and the frame S', in S, is $u_B-v_B$, the speed of the object, as observed in S',
      is the transformed value of this relative velocity. If the origins of S and S' and the moving
      object all have the same $x$-coordinate at time $t_B = 0$, and $u _B > v_B$, the separation,
     $\Delta x_B$, in the frame S, of the object from the origin of S' at time $t_B$ is 
 \begin{equation}
    \Delta x_B = (u_B-v_B)t_B.
  \end{equation}
     If $\bar{u}'_T$ is the velocity of the object in S', in the positive $x'$ direction, in the primary experiment, the separation of the
    object from the origin
      of S' at time $t'_B$ is
\begin{equation}
    \Delta x'_T = \bar{u}'_T t'_B.
  \end{equation}
  The Lorentz invariance, (4.10), of the spatial separation of the object from the origin of S',
    at the corresponding times $t_B$ and $t'_B$, which implies, for  $u _B > v_B$,
   $\Delta x_B = \Delta x'_T$,  and the TD relation: $t_B = \gamma_B t'_T$ (c.f. Eq.~(3.9)) then gives the 
   transformation law for the relative velocity of the object and S' as:
 \begin{equation}
  \bar{u}'_T = \gamma_B(u_B-v_B)
  \end{equation}
 where  $\gamma_B \equiv 1/\sqrt{1-(v_B/c)^2}$. 
  As above, the bar in the symbol $\bar{u}'_T$ indicates that it is a calculated
  quantity as contrasted with the 
  assumed initial values, in this case, of $v_B$ and $u_B$.
 In the special case $u_B = 0$, (5.19) is the transformation law of the relative velocity of S and S' between the base 
  frame S and the travelling frame S':
 \begin{equation}
   -\bar{u}'_T \equiv \bar{v}'_T =  \gamma_Bv_B
  \end{equation}
  where $\bar{v}'_T$ is defined as the velocity of S relative to S', in S', in the direction of the negative
   $x'$-axis in the primary experiment. Thus the Reciprocity Principle (RP)~\cite{BG}, that \newline
   `` the velocity of an inertial frame of reference 
   S', with respect to another inertial frame of reference S, is equal and opposite to velocity of S relative
   to S' '', although true in Galilean relativity, no longer holds in special relativity, being replaced 
   by the reciprocity relation (5.20), when the space-time LT is used to transform
    events, in a particular space-time experiment, from one frame into another. As derivation of the latter
    equation shows, the breakdown of the
    RP is a necessary consequence of the definition of a relative velocity, the invariance of spatial
    intervals, and TD. The reciprocity relation for an experiment with a reciprocal configuration where
    S' is the base frame and S the travelling frame is   
 \begin{equation}
  \bar{v}_T =  \gamma'_B v'_B
  \end{equation}
   where $\gamma'_B \equiv 1/\sqrt{1-(v'_B/c)^2}$. Eq.~(5.21) is obtained from (5.20) by exchange of primed and
   unprimed quantities. Reciprocal experiments correspond to the special case where $v'_B = v_B \equiv v$.
   \par Thus, in special relativity, the RP should be replaced a `Kinematical Reciprocity Principle' (KRP)~\cite{JHFSTP3}:
   `the velocity of an inertial frame of reference S' relative to another such frame S in a space-time experiment
     is equal and opposite to the velocity of S relative to S' in the reciprocal experiment'. This
    statement, which is actually the definition of a reciprocal experiment, rather than a relation
       between velocities in different frames in the same space-time experiment, is applicable in both 
    special and Galilean relativity.

 \SECTION{\bf{Einstein's train-embankement thought experiment}}
    A straightforward application of the relative velocity transformation law (5.19) is to the analysis
   of the much-discussed train-embankment thought experiment\cite{EinTETE}. This was introduced by Einstein
   in the popular book `Relativity, the Special and General Theory' with the intention to illustrate, in a 
   simple way, `relativity of simultaneity'. Light signals are produced by lightning strikes which simultaneously
   hit an embankment 
     at positions coincident with the front and back ends of a moving train. The signals are
     seen by an observer, O$_T$, at the middle of the train and an observer,  O$_B$, on the
     embankment, aligned with
     O$_T$ at the instant of the lightning strikes. The light signals are observed simultaneously
     by  O$_B$ who concludes that the lightning strikes are simultaneous. Because of the
     relative motion of O$_T$ and the light signals, the latter are not observed by O$_T$
     at the same time. Invoking the constancy of the speed of light in the train frame, Einstein
     concludes that O$_T$ would not judge the strikes to be simultaneous, giving rise to a
     `relativity of simultaneity' effect between the train and embankment frames.

\begin{figure}[htbp]
\begin{center}\hspace*{-0.5cm}\mbox{
\epsfysize10.0cm\epsffile{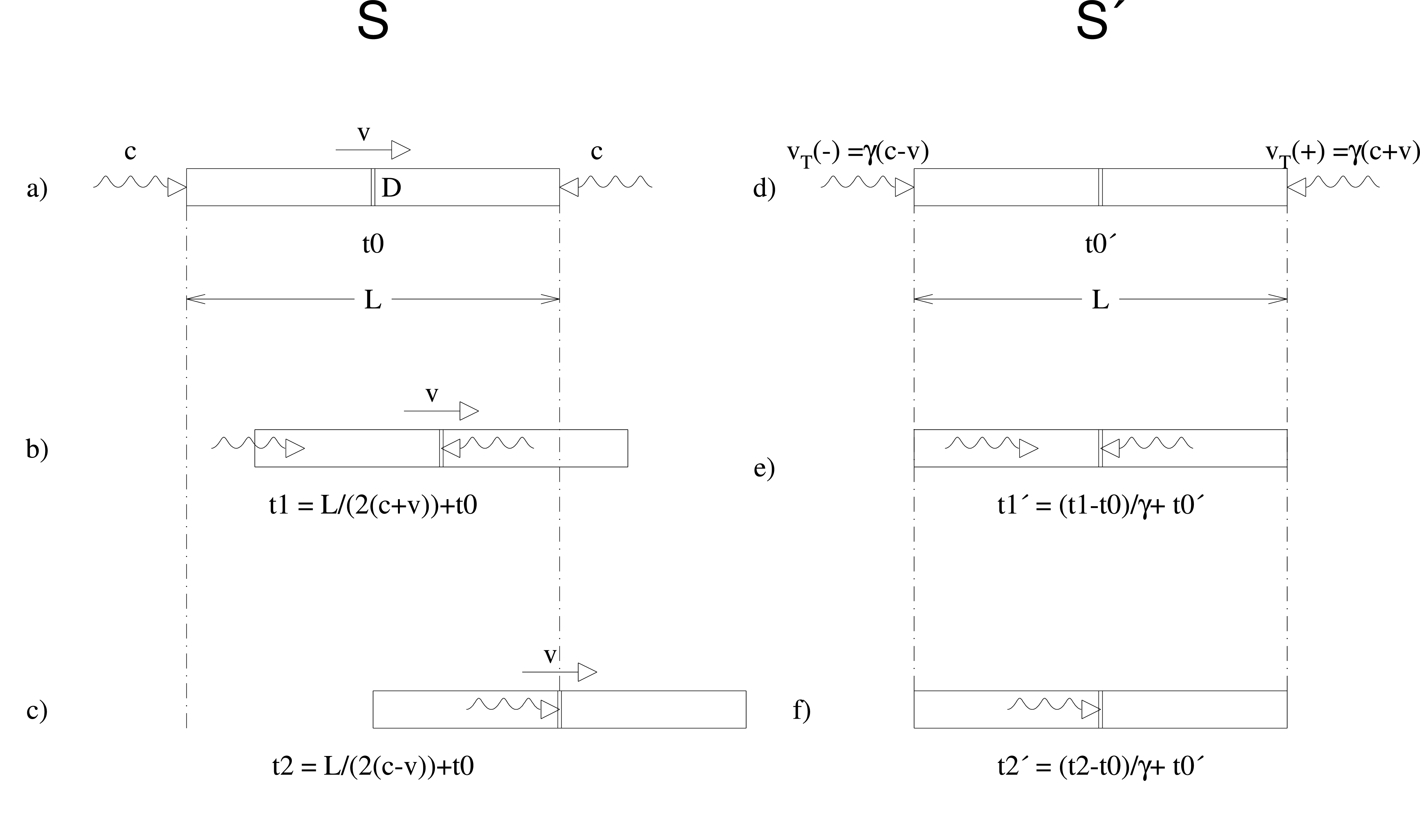}}
\caption{{\em Analysis of Einstein's train-embankment thought experiment.
   Configurations a),b) and c) in the embankment frame (S); d), e) and f) in the train frame (S').
      $v = c/2$, $\gamma = 2/\sqrt{3}$.  See text for discussion.}}
\label{fig-fig1}
\end{center}
\end{figure}

       \par This train-embankment thought experiment (TETE) is now analysed in terms of
       the concepts and nomenclature introduced above\footnote{A similar analysis is presented
        in Ref.~\cite{JHFSTP3}.}. The observer O$_T$
       is replaced by a two-sided light detector, D, at the middle of the train. The
       latter moves to the right with speed $v$. The embankment frame, S, is the base 
        frame of the experiment, the train frame, S', is the travelling frame.
    At time $t0$ in S (Fig. 1a) light signals moving at speed $c$ in the embankment frame
     are emitted, and move towards D. The light signals are also `travelling objects' in the source frame
     S. The essential input parameters of the problem, $v$ and $c$ are therefore fixed in the frame
     S. In accordance with Eq.~(4.13) the length of the train, $L$, is invariant. At time in S
      $t1 = L/[2(c+v)] + t0$ (Fig. 4b) the left-moving light signal strikes D, and at time in S
   $t2 = L/[2(c-v)] + t0$.(Fig. 1c) the right-moving light signal strikes D. The configurations
     in S' corresponding to those in S in Figs. 1a,b,c are shown in  Figs. 1d,e,f respectively.
     The velocity transformation formula (5.19) implies that the speed in S' of the
    right-moving light signal relative to D is $\bar{v}_T(-) = \gamma(c-v)$ while that of the
 left-moving light signal is $\bar{v}_T(+) = \gamma(c+v)$. The pattern of detection events in S and S' is
    then the same, the only difference being that that the velocities of the light
    signals relative to D are greater in S' by the factor $\gamma$ ---a necessary consequence of
    time dilation and the invariance of length intervals. The left-moving light signal
    is then observed in S' at the time:
   \begin{equation}
      t1' = \frac{L}{2 \gamma(c+v)} +t0' = \frac{t1-t0}{\gamma} + t0' 
    \end{equation}
     and the right-moving one at the time:
   \begin{equation}
      t2' = \frac{L}{2 \gamma(c-v)} +t0' = \frac{t2-t0}{\gamma} + t0'. 
    \end{equation}
     The time dilation  effect for the travelling frame S' is manifest in these equations.
     It is seen to be a consequence of the relative velocity transformation formula (5.19), not of LC.
     \par On the assumption that an experimenter analysing the signals received by D knows
      the essential parameters of the problem, $L$, $v$, and $c$, the measured times 
     $t1'$ and $t2'$ in the train frame can be used to decide whether the 
       left and right moving light signals were emitted simultaneously in this frame or not.
        If the right-moving and left-moving signals are emitted at times $t0'(-)$ and
        $t0'(+)$ respectively then (6.1) and (6.2) are modified to:
   \begin{equation}
      t1' = \frac{L}{2 \gamma(c+v)} +t0'(+) = \frac{t1-t0}{\gamma} + t0'(+) 
    \end{equation}
     and 
   \begin{equation}
      t2' = \frac{L}{2 \gamma(c-v)} +t0'(-) = \frac{t2-t0}{\gamma} + t0'(-). 
    \end{equation}
     Subtracting (6.3) from (6.4) and rearranging:
   \begin{equation}
     t2' -  t1' =  t0'(-)- t0'(+) +\frac{\gamma \beta L}{c}.
   \end{equation}
   The observed time difference $t2' -  t1'$ and knowledge of the value
   of $\gamma \beta L/c$ then enables determination of $t0'(-)- t0'(+)$ so that
    the simultaneity of emission of the light signals can be tested. For the event configurations
     shown in Fig. 1 it would be indeed concluded that $t0'(-)= t0'(+)$, 
      so the emission of the signals is found to be simultaneous in the train frame,
      contrary to Einstein's assertion in Ref.~\cite{EinTETE}. The essential flaw in Einstein's
    argument was the failure to distinguish between the speed of light, relative to some fixed
      object in an inertial frame, and the speed of light relative to some moving object
     in the same frame, which is what is relevant for the analysis of the TETE.
      Einstein's interpretation corresponds to replacing  $t2'$ and  $t1'$
      by $t2$ and $t1$, so that only events in the embankment frame are 
      considered, and making the replacements, (confusing the speed of light in an inertial frame, with the relative
      speed of light and a moving object in the same frame): $\gamma(c \pm v) \rightarrow c$ in (6.1) and (6.2)
      giving:
    \begin{equation}
       t0'(-)- t0'(+) = t2  - t1 = \frac{\gamma^2 \beta L}{c}.
    \end{equation}
       This leads to Einstein's false conclusion that the light signal emission
       events would be found to be non-simultaneous in the train frame.
     \par An analysis of the TETE in a previous paper~\cite{JHFTETE} by
      the present author also concluded that the train observer would judge the lightning
      strikes to be simulataneous, but the reasoning leading to this conclusion was
      fallacious. At the time of writing Ref.~\cite{JHFTETE} I had not understood correctly
      the distinction between an experiment and its reciprocal and the difference in physical
     interpretation of a space-time and a kinematial LT explained in Sections 3 and 5 above.
      I incorrectly assumed that the kinematical LT relating two base frame velocities
      was valid in a single space-time experiment, that is, for transformation between a base 
       frame and a travelling frame. Thus the velocities of both photons in S' in Fig. 1
      of the present paper were assumed to be $c$. Since the lightning strikes are (see Section 3 above)
      simultaneous in both S and S' the train observer would then see the light signals they emit
       at the same time and conclude that the strikes are simultaneous. This is a correct description,
       in the train frame,
       of the physically independent experiment that is reciprocal to the one proposed by Einstein (i.e. the one
       where S' is the the base, not the travelling frame)
       not the correct description, in the train frame, of Einstein's experiment. The mistake
       in Ref.~\cite{JHFTETE} was the hitherto universal, but erroneous,
       assumption that events defined in the base frames
       of an experiment and its reciprocal are related by the space-time LT.
        \par As demonstrated by Post~\cite{Post},
    the different relative velocities of the light signals and the 
        detector D, shown in Fig.~1 is also the physical basis of the Sagnac effect~\cite{Sagnac,MG}, where
         light signals move with different vlocities relative to a rotating interferometer.

 \SECTION{\bf{Sartori's two-train thought experiment}}
  Sartori~\cite{Sartori} proposed the thought experiment shown in Fig. 2. In the base frame S,
  the rest frame of the platform P, two trains T1 and T2, with proper frames S' and S'' respectively, 
   move to the right with speeds $v$
  and $u$ respectively, where $u > v$. These base frame velocities are the fixed input parameters of the problem. As in the
  previous Section, to lighten the notation, the base frame labels on these quantities are omitted.
  Initially (Fig. 2a) when $t = t1 = 0$ , T1 is aligned with P and distant L1 from T2. 
  At time $t = t2 = L1/u$, T2 is aligned with P and T1 is distant L2 from P (Fig. 1b). Since $u>v$, T1
  is aligned with T2 at time $t = t3 = L1/(u-v)$ when T1 and T2 are distant L3 from P (Fig. 2c).
  The corresponding configurations as observed in the travelling frames S' and S'' are shown in Figs. 3 and 4.
  Because of the invariance of spatial separations the corresponding spatial configurations are identical
 to those of Fig. 2, whereas the times of the corresponding events are scaled by the TD factors
   $1/\gamma_v$,  $1/\gamma_u$ respectively. The travelling frame
   velocities as given by the relative velocity transformation formula (5.19) are  shown in Figs. 3 and 4.

\begin{figure}[htbp]
\begin{center}\hspace*{-0.5cm}\mbox{
\epsfysize10.0cm\epsffile{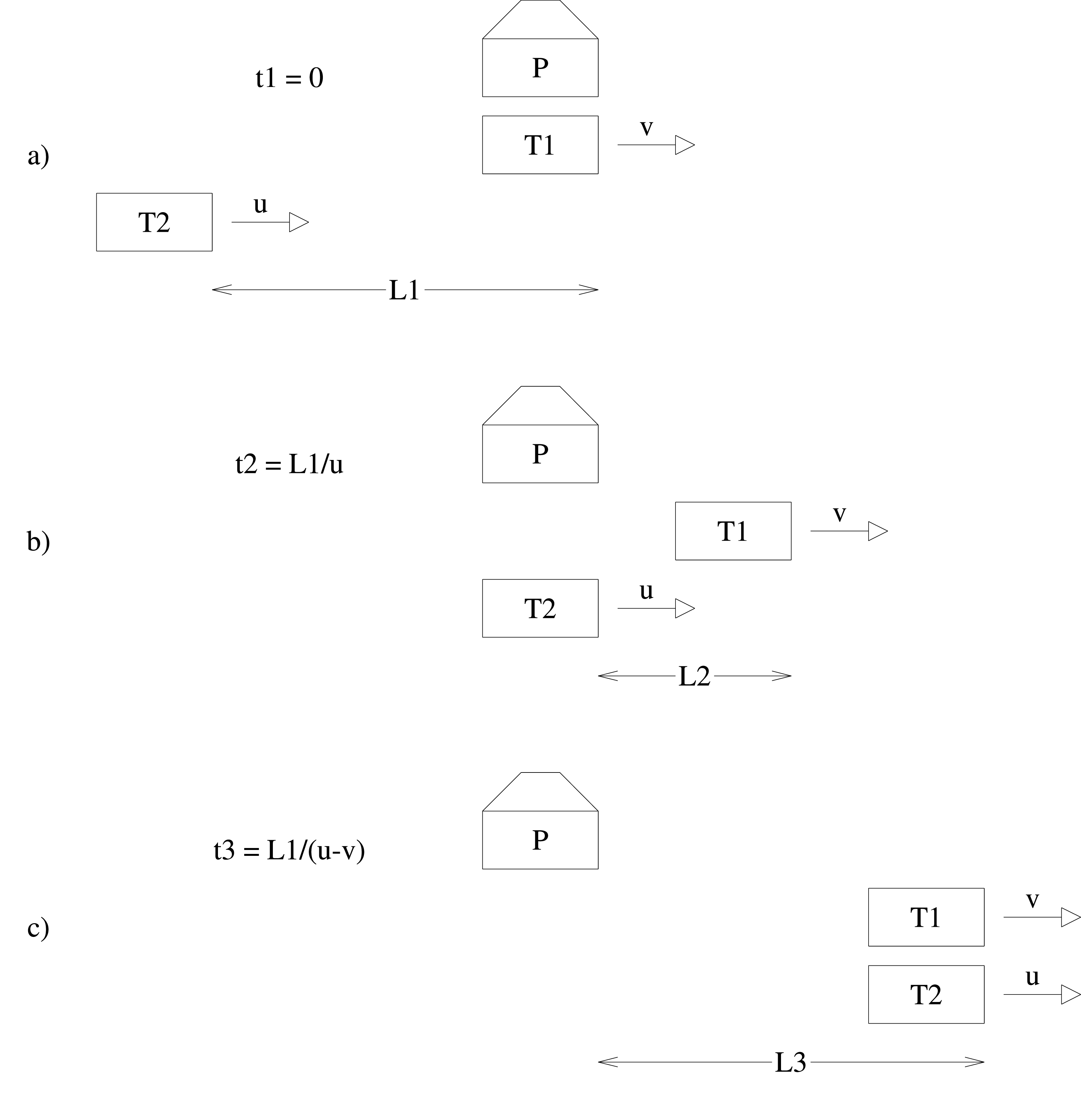}}
\caption{{\em Spatial coincidence events as observed in the base frame S (the
 rest frame of P) in which the velocities $v = 0.4c$ and $u = 0.8c$ of T1 and T2,
  respectively, are specified.
  a) Event1, T1 opposite P,  b) Event2, T2 opposite P,  c) Event3, T1 opposite T2.}}
\label{fig-fig2}
\end{center}
\end{figure}

\begin{figure}[htbp]
\begin{center}\hspace*{-0.5cm}\mbox{
\epsfysize10.0cm\epsffile{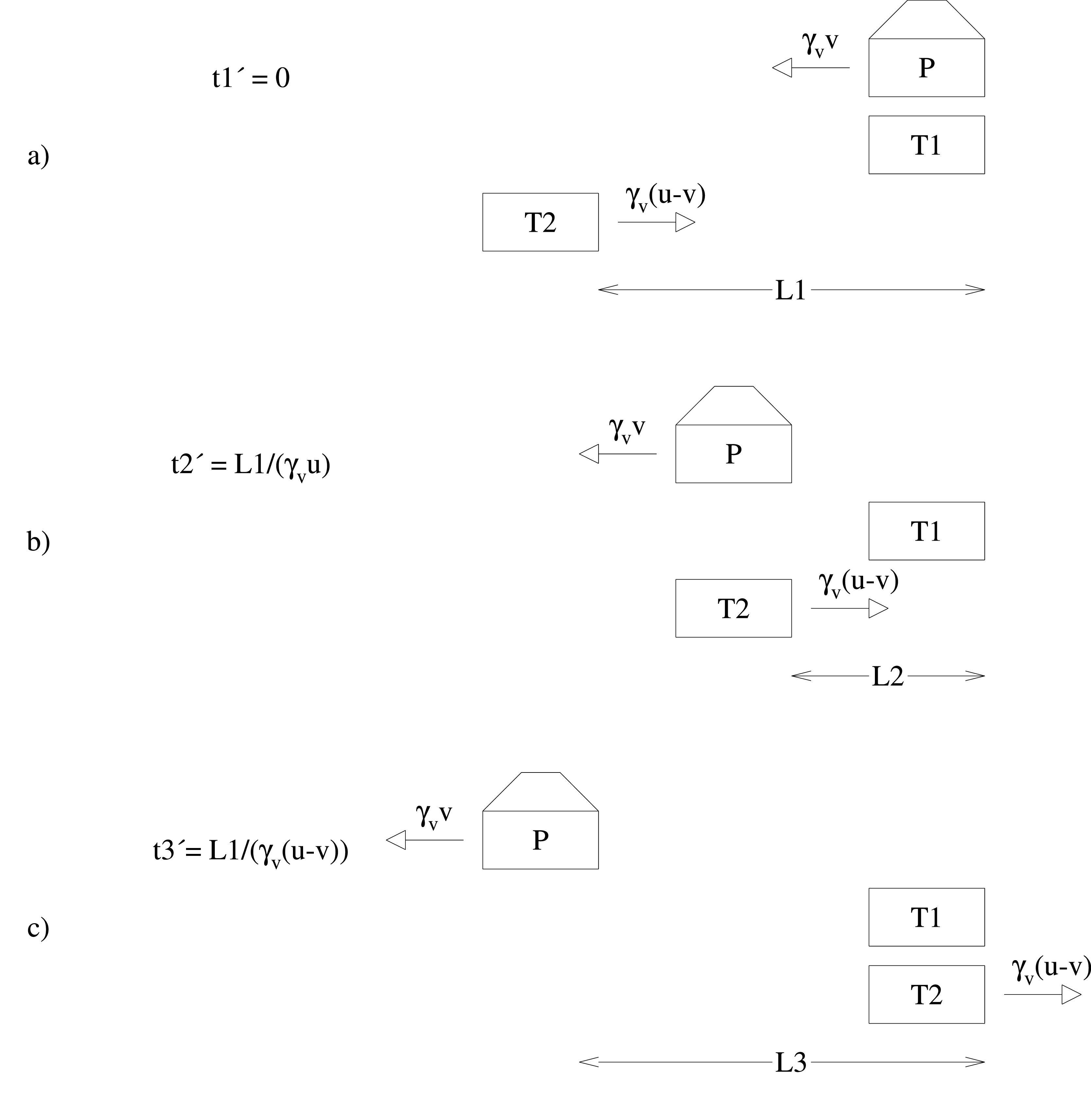}}
\caption{{\em Spatial coincidence events as observed in the travelling frame S' (the
 rest frame of T1). a) Event1, T1 opposite P,  b) Event2, T2 opposite P, 
 c) Event3, T1 opposite T2.}}
\label{fig-fig3}
\end{center}
\end{figure}

\begin{figure}[htbp]
\begin{center}\hspace*{-0.5cm}\mbox{
\epsfysize10.0cm\epsffile{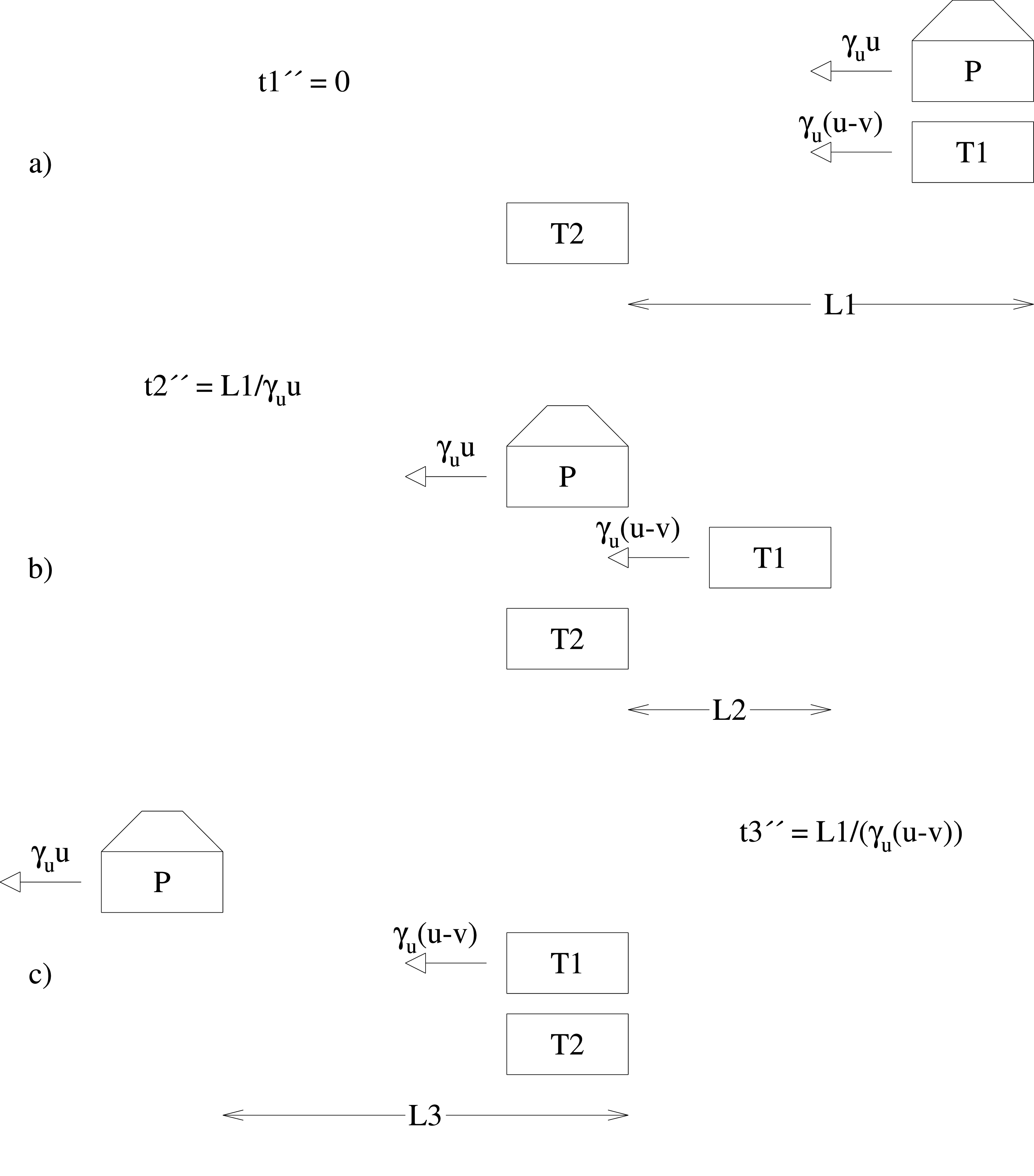}}
\caption{{\em  Spatial coincidence events as observed in the travelling frame rest S''
  (the rest frame of T2). a) Event1, T1 opposite P,  b) Event2, T2 opposite P, 
 c) Event3, T1 opposite T2.}}
\label{fig-fig4}
\end{center}
\end{figure}

\begin{table}
\begin{center}
\begin{tabular}{|c|c|c|c|c|} \hline  
B & T & $v$(P)~~~~$v$(T1)~~~~$v$(T2) & $v'$(P)~~~~$v'$(T1)~~~~$v'$(T2) & $v''$(P)~~~~$v''$(T1)~~~~$v''$(T2)  \\
 \hline 
& & & & \\
 S  & S',S''& 0~~~~~~~~~~~~$v$~~~~~~~~~~$u$ & $-\gamma_v v$~~~~~~0~~~~$\gamma_v(u-v)$ &  $-\gamma_u u$~~~~$-\gamma_u(u-v)$ ~~0 \\
 & & & & \\
 \hline
& & & &  \\
S'  & S,S''& 0~~~~~ $\gamma_v v$~~$\gamma_v(v+w)$ & $-v$~~~~~~~~0~~~~~~~~~$w$ &
   $-\gamma_w(v+w)$~$-\gamma_w w$~~~~0 \\
 & & & &  \\
 \hline
& & & &  \\
S''  & S,S''& 0~~~~~$\gamma_u(u-w)$~~~~$\gamma_u u$ &  $-\gamma_w(u-w)$~~~0~~~~~~$\gamma_w w$ &
   $-u$~~~~~~~~$-w$~~~~~~~~~~~~0 \\
 & & & &  \\
 \hline
\end{tabular}
\caption[]{{\em Base frame (B) and travelling frame (T) velocities in various frames.
 The base frame velocities are related by the parallel velocity addition formula 
 (5.16) while the travelling frame velocities are derived from base frame velocities
  using the relative velocity transformation formula (5.19). Each row of velocities
  specifies a physically-independent space-time experiment. $w = (u-v)/[1-(uv)/c^2]$.  The experiment shown
   in Figs.2-4 is that shown in the first row.}}
\end{center}
\end{table}


\begin{table}
\begin{center}
\begin{tabular}{|c|c|cc|cc|cc|cc|} \hline  
B & T & $t_2$ & $t_3$ & $t'_2$ & $t'_3$ & $t''_2$ & $t''_3$ & $L2$ & $L3$  \\
 \hline 
& & & & & & & & & \\
 S  & S',S''&  $\frac{L1}{u}$ &  $\frac{L1}{u-v}$ &  $\frac{L1}{\gamma_v u}$ &
   $\frac{L1}{\gamma_v(u-v)}$ &  $\frac{L1}{\gamma_u u}$ & $\frac{L1}{\gamma_u(u-v)}$
   & $\frac{v L1}{u}$ & $\frac{v L1}{u-v}$ \\
 & & & & & & & & &  \\
 \hline
& & & & & & & & &  \\
 S' & S, S''&  $\frac{L1}{\gamma_v(w+v)}$ &  $\frac{L1}{\gamma_vw}$ &  $\frac{L1}{w+v}$ &
   $\frac{L1}{w}$ &  $\frac{L1}{\gamma_w(w+v)}$ & $\frac{L1}{\gamma_ww}$
   & $\frac{v L1}{(w+v)}$ & $\frac{vL1}{w}$ \\
 & & & & & & & & &  \\
 \hline
& & & & & & & & &  \\
S'' & S, S'&  $\frac{L1}{\gamma_u u}$ &  $\frac{L1}{\gamma_u w}$ &  $\frac{L1}{\gamma_w u}$ &
   $\frac{L1}{\gamma_w w}$ &  $\frac{L1}{u}$ & $\frac{L1}{w}$
   & $\frac{L1(u-w)}{u}$ & $\frac{(u-w) L1}{w}$ \\
 & & & & & & & & &  \\
 \hline
\end{tabular}
\caption[]{{\em Times and spatial separations of the coincidence events 2 and 3 in
  different frames for the three space-time experiments specified in the rows 
  of Table 1.}}      
\end{center}
\end{table}

 \par The configurations shown in Figs. 2-4 correspond to $u = 0.8 c$, $v = 0.4c$. 
   By performing a kinematical transformation according to Eq.~s(5.8)-(5.10)
   or (5.11)-(5.13), between S and S' or S and S'' configurations may be obtained in which T1  is 
   at rest (configuration in S') or T2 at rest (configuration in S''). In these cases the parallel
   base frame velocity transformation (5.16) is applicable. In S' P has velocity -$v$ and T2 velocity \newline
 $w =(u-v)/(1-uv/c^2)$, while in S'' P has velocity -$u$ and T1 velocity -$w$. These configurations
  constitute base frame parameters for two independent space-time experiments each of which are reciprocal 
  to the primary experiment shown in Figs.2-4. In the first reciprocal experiment S' is the base frame and
   S and S'' are travelling frames while in the second S'' is the base frame and S and S' are the travelling frames. 
   The travelling  and base frame velocities for all three experiments are presented in Table 1;
   the base frame velocities transform according to Eq.~(5.16), while the travelling frame velocities
   are calculated using the relative velocity transformation formula (5.19). In Table 2 are presented, for the 
   primary experiment and the two reciprocal experiments,
   the values of $t_2$, $t'_2$ and $t''_2$ (when P and T2 are aligned) and
    $t_3$, $t'_3$ and $t''_3$ (when T1 and T2 are aligned) in the frames S, S' and S'', respectively,
    as well as the (invariant) separations: L2, of P and T1 at the instant of P2-T2 alignment and
    L3 of P and T1 at the instant of T1-T2 alignment. The different values of the times and separations
    in the primary and two reciprocal experiments make manifest the physical independence of these
    experiments even though their kinematical configurations are related by the kinematical LT
    of Eq.~s(5.8)-(5.10). The initial spatial separation of P and T1, $L1$, is the same in all
    three experiments. Several different TD relations can be read off from the entries of Table 2:
    \begin{eqnarray}
  {\rm S(B),~S'(T),~S''(T)}:~~~~t({\rm B}) & = & \gamma_vt'({\rm T}),~~~~t({\rm B}) = \gamma_u t''({\rm T}), \\
  {\rm S'(B),~S(T),~S''(T)}:~~~~t'({\rm B}) & = & \gamma_vt({\rm T}),~~~~t'({\rm B}) = \gamma_w t''({\rm T}), \\
  {\rm S''(B),~S(T),~S'(T)}:~~~~t''({\rm B}) & = & \gamma_ut({\rm T}),~~~~t''({\rm B}) = \gamma_w t'({\rm T}).
     \end{eqnarray}
    It is interesting to compare these predictions with assumption made by Sartori in the paper 
    where the thought experiment shown in Figs. 1-3 was first proposed~\cite{Sartori}. The aim of this
    paper was to present a simple derivation of the parallel velocity addition formula (5.16) without
    direct use of the LT. In the paper it was claimed to derive (5.16) taking as initial postulate
    the TD effect. The first incorrect assumption of Ref.~\cite{Sartori} was that the base frame
    configurations of the primary experiment\footnote{The symbols $[{\rm S(B),~S'(T)}]$
    and  $[{\rm S'(B),~S(T)}]$ specify, in an evident notation, an experiment and its (physically 
     independent) reciprocal.} $[{\rm S(B),~S'(T)}]$ and the reciprocal experiment
      $[{\rm S'(B),~S(T)}]$, which are indeed related by a kinematical LT, are also related by a 
      space-time LT. Thus it was assumed that in the relations\footnote{Eq.~s.(7.4) and (7.5) correspond to
        Eq.~s.(2) and (8), respectively, of Ref.~\cite{Sartori}.} that may be derived from the
      entries of Table 2:
    \begin{eqnarray}
 t_3({\rm B})  & = & \frac{u t_2({\rm B})}{u-v}, \\
  t'_3({\rm B})  & = & \frac{(w+v)t'_2({\rm B})}{w}
  \end{eqnarray}
 that $t_2({\rm B})$ and $t'_2({\rm B})$ and  $t_3({\rm B})$ and $t'_3({\rm B})$ are related by TD relations:
    \begin{eqnarray}
 t'_2({\rm B})  & = &  \gamma_v t_2({\rm B}), \\
  t_3({\rm B})  & = &   \gamma_v t'_3({\rm B}).
  \end{eqnarray}
  Taking the ratio of (7.4) to (7.5) and using (7.6) and (7.7) to eliminate the ratios
  $t_2({\rm B})/ t'_2({\rm B})$ and $t_3({\rm B})/ t'_3({\rm B})$ from the resulting equation gives:
    \begin{equation}
     \gamma_v^2(u-v)(w+v) = uw
   \end{equation}
      which when solved for $w$ in terms of $v$ and $u$ yields the parallel velocity addition
      formula (5.16). 
  \par The following comments may be made on this calculation:
   \begin{itemize}
    \item[(i)] The base frame configurations of the independent experiments $[{\rm S(B),~S'(T)}]$
       and   $[{\rm S'(B),~S(T)}]$ although related by the kinematical LT of Eq.~s(5.8)-(5.9) are not connected
      by the space-time LT ---for  $[{\rm S(B),~S'(T)}]$ the speed of P in the frame S' is (see Fig. 3a)
      $\gamma_v v$ not $v$.
    \item[(ii)] Comparison of (7.6) and (7.7) with (7.1) and (7.2) above shows that the former
      formulas are inconsistent. For Event 2 the formula (7.6) corresponds to the TD effect
      for the experiment  $[{\rm S'(B),~S(T)}]$, as in (7.2), whereas for Event 3 the formula (7.7)
      corresponds ot the independent experiment  $[{\rm S(B),~S'(T)}]$ as in (7.1). It is clear that,
      for example, the times $t'_2$ and $t'_3$ must both be times of clocks at rest in S' but
      observed in motion from S in the primary experiment shown in Figs 1-3, $[{\rm S(B),~S'(T)}]$,
      and times of clocks at rest in S' and observed in the same frame for the reciprocal
      experiment  $[{\rm S'(B),~S(T)}]$.
    \end{itemize}
     \par The argument given by Sartori for (7.6) and (7.7) is that, for (7.6) `the Events 1 and 2
      occur at the same position in S' so that $t_2$ is a proper time interval', and for (7.7)
      that `the Events 1 and 3 occur at the same position in S' so that $t'_3$ is a
      proper time interval'. Actually all the times in the TD relations shown in (7.1)-(7.3)
      are `proper time intervals' recorded by some clock. Given the existence of an array of
       synchonised clocks in each frame, it is of no importance, for the timing of
      two events whether, or not, their times are both measured locally by the same clock.
      Suppose that in Fig. 3 there is a clock C' at rest in S', synchronised with a clock at T1,
     distant L2 from it, such that Event 2 in Fig. 3b is local at C'. Because C' and a synchronised
      clock at T1 both record $t'= 0$  at the epoch of Event 1, the time interval
    in S', between events 1 and 2 in S',  is correctly given by the epoch $t'_2$, as measured by
    C', of the Event 2 which 
    is local at C'. Since  $t'_2$ is a`proper time interval' measured by the clock
     C', then if $t_3 = \gamma_v t'_3$ as in (7.6), as measured by a local clock at T1, then also  $t_2 = \gamma_v t'_2$
      as measured by the local clock  C', in contradiction with Sartori's assumption (7.6).
     \par Summarising, Sartori's analysis assumes incorrectly that the base frames of an
      experiment and its reciprocal are related by the space-time LT and uses TD relations
     in an inconsistent manner, (7.7) being applicable to the primary experiment shown in
     Figs 2-4:  $[{\rm S(B),~S'(T)}]$, and (7.6) to the reciprocal experiment: $[{\rm S'(B),~S(T)}]$.
       That the algebraic manipulation of (7.4)-(7.7) yields the correct velocity 
      addition formula (5.16) must then be considered as purely fortuitous.
      \par In my previous analysis~\cite{JHFSartori} of Sartori's thought experiment, the same 
         mistake of principle was made as in that of the train/embankment experiment in 
         Ref.~\cite{JHFTETE}. In common with Sartori, it was assumed that the the kinematical
       configurations of a primary experiment and its reciprocal correspond to the base and travelling
       frame configurations of the primary experiment. That is (see Figs. 1 and 2 of Ref.~\cite{Sartori})
       that the velocities in the frame S' in the experiment shown in Figs. 2-4 and the first row
       of Table 1 of the present paper, were given instead by the base frame velocities in S' of the 
       reciprocal experiment shown in the second row of Table 1. Thus, it is falsely assumed that
      the base frame configurations of an experiment and its reciprocal are actually the base and
      travelling frame configurations of the primary experiment, and that events in these
      two frames are connected by the space-time LT. This leads to false predictions~\cite{JHFSartori} of
      the breakdown of the Lorentz invariance of spatial intervals in different inertial
       frames and ratios between time intervals observed in different inertial frames differing
       from the TD effect.


\pagebreak

\end{document}